\documentclass[10pt]{iopart}

\begin{document}

\def\Dirac{i\partial\!\!\!\!/}
\def\DDirac{iD\!\!\!\!/}
\def\dirac{i\partial\!\!\!\!/-eA\!\!\!\!/}

\newcommand{\gc}{\gamma _M^0}
\newcommand{\gu}{\gamma _M^1}
\newcommand{\gd}{\gamma _M^2}
\newcommand{\gce}{\gamma _0}
\newcommand{\gue}{\gamma _1}
\newcommand{\beq}{\begin{eqnarray}}
\newcommand{\eeq}{\end{eqnarray}}
\newcommand{\nn}{\nonumber}

\makeatletter
\renewcommand{\theenumi}{\arabic{enumi}}
\article[Relativistic Landau problem]{\fl Seventh International
Workshop Quantum Field Theory under the Influence of External
Conditions, QFEXT'05, Barcelona, Spain}{Relativistic Landau
problem at finite temperature}
\author{C G Beneventano\footnote{Member of CONICET} and
E M Santangelo\footnote{Member of CONICET}}
\address{Departamento de F{\'\i}sica, Universidad Nacional de La Plata
\\Instituto de F{\'\i}sica de La Plata, UNLP-CONICET\\
C.C. 67, 1900 La Plata, Argentina}
\ead{\mailto{gabriela@obelix.fisica.unlp.edu.ar},
\mailto{mariel@obelix.fisica.unlp.edu.ar}}
\begin{abstract}
We study the zero temperature Casimir energy and fermion number
for Dirac fields in a 2+1-dimensional Minkowski space-time,in the
presence of a uniform magnetic field perpendicular to the spatial
manifold. Then, we go to the finite-temperature problem, with a
chemical potential, introduced as a uniform zero component of
the gauge potential. By performing a Lorentz boost, we obtain Hall's conductivity in the case of crossed electric and magnetic fields. \\[.3cm]
{\bf Subject Classification} 

\end{abstract}
\section{Introduction}\label{intr}

The quantization of the Hall conductivity \cite{klitzing} is a remarkable quantum phenomenon,
which occurs in two-dimensional electron systems, at low temperatures and strong perpendicular
magnetic fields. Most proposed explanations for this phenomenon \cite{basics} rely on
Schr{\"o}edinger's one-particle theory and the introduction of a filling fraction of the Landau
levels, which must be assumed to be integer to reproduce the observed behavior.

It is the aim of this paper to show that, in the context of relativistic field theory, such
behavior naturally arises, as a consequence of the spin-statistics theorem.

In section \ref{Sect1} we present the theory of Dirac fields in
$2+1$ Minkowski space-time, interacting with a magnetic background
field perpendicular to the spatial plane, and evaluate the vaccum
expectation values of the energy and fermion density. Section
\ref{Sect2} contains the generalities of the same theory in
Euclidean $3$-dimensional space, and presents the eigenvalues of
the corresponding Dirac operator. From such eigenvalues, the
partition function is evaluated in section \ref{Sect3}. Section
\ref{Sect4} contains the resulting free energy and mean particle
density at finite temperature. Finally, in section \ref{Sect5} we
perform an adequate Lorentz boost in order to consider the problem
of fermions interaction with crossed electric and magnetic fields,
and obtain the Hall conductivity.

\section{Zero-temperature problem}\label{Sect1}
We study a $2+1$-dimensional theory of Dirac fields, in the
presence of a uniform background magnetic field perpendicular to
the spatial plane. We choose the metric $(-,+,+)\,,$ natural units
$\hbar=c=1$, and adopt the following representation for the Dirac
matrices: $ \gc=i\sigma_3,\,\gu=\sigma_2\, {\rm and}\,
\gd=\sigma_1$.

The Hamiltonian can be determined from the
solutions of the Dirac equation $(\dirac)\Psi=0$, where $-e$ is the negative charge of the electron. In the Landau gauge $A=(0,0,Bx)\,,$
with $B>0$. Thus, after setting
$\Psi(t,x,y)=e^{-iEt}\psi(x,y)$, we get the Hamiltonian $H=i\sigma_1
\partial_x-i\sigma_2
\partial_y+\sigma_2 eB x$.

In order to solve the eigenvalue problem for this Hamiltonian, we
take \beq \psi_k(x,y)=\left(\begin{array}{c}
  \varphi_k(x,y) \\
  \chi_k(x,y) \\
\end{array}\right) =\frac{1}{\sqrt{2\pi}}
\left(\begin{array}{c}
e^{iky} \varphi_{k} (x)\\
e^{iky} \chi_{k} (x) \\
\end{array}\right)\,.\eeq

This leads to the following system of first order equations \beq
\nn (i\partial_x -i k -i e B
x)\chi_{k}&=&E\varphi_{k}\\(i\partial_x +i k + i e B
x)\varphi_{k}&=& E\chi_{k} \,.\eeq

After imposing that the eigenfunctions be well-behaved for $x
\rightarrow \pm \infty$, we find two types of solutions to our
problem \footnote{For the other non-equivalent representation of
the gamma matrices in $2+1$ dimensions, the spectrum doesn't
change.}:

1. A definite-chirality zero mode ($E_0=0$)

2. A set of eigenfunctions corresponding to the symmetric spectrum
\hfil\break $E_n=\pm \sqrt{2n e B}\,,\, n=1,..., \infty$.

In all cases, the eigenfunctions can be written in terms of
Hermite polynomials, and all eigenvalues exhibit the well known
Landau degeneracy per unit area:\beq \Delta_L\ = \frac{e
B}{2\pi}\,. \label{deg}\eeq

The vacuum expectation value of the energy per unit area, defined
through a zeta function regularization (see, for example,
\cite{elizalde} and references therein), is given by $  E_C =
-\left.\frac{\Delta_L}{2} \sum_{E_n\neq 0}
|E_n|^{-s}\right\rfloor_{s=-1}$.

In the present case, we have ($\alpha$ is an arbitrary parameter
with mass dimension, introduced to render the complex powers
dimensionless) \beq\fl E_C(B) = -\left.\frac{\Delta_L\alpha}{2} 2
\sum_{n=1}^{\infty} \left(\frac{\sqrt{2n e
B}}{\alpha}\right)^{-s}\right\rfloor_{s=-1}= -\Delta_L \sqrt{2 e
B}\, \zeta_R \left(-\frac{1}{2}\right)\,.\eeq

Always in the zeta-function regularization framework, the fermion
number is \cite{ns} \beq \nn N(B) &=& -\left.\frac{\Delta_L}{2}
\left(\sum_{E_n > 0} |E_n|^{-s}-\sum_{E_n < 0}
|E_n|^{-s}\right)\right\rfloor_{s=0}+N_0\,,\label{numero}\eeq
where $N_0$ is the contribution coming from zero modes.

In our case, the nonvanishing spectrum is symmetric. So, only the
zero mode, which is charge self-conjugate, contributes. This gives
as a result \cite{ns}
 \beq N(B) =\pm \frac{\Delta_L}{2}\,.\label{nf}\eeq
Or, equivalently, for the vacuum expectation value of the charge
density \beq j^0(B) =\mp e\frac{\Delta_L}{2}\,.\label{jcero}\eeq

\section{The theory at finite temperature with chemical potential}\label{Sect2}

In order to study the effect of temperature, we go to Euclidean
space, with the metric $(+,+,+)\,.$ To this end, we take the
Euclidean gamma matrices to be $\gce=i\gc=-\sigma_3$,
$\gue=\gu=\sigma_2$, $\gamma_2=\gamma_M^2=\sigma_1$. We will
follow \cite{actor} in introducing the chemical potential as an
imaginary $A_0=-i\frac{\mu}{e}$ in Euclidean space. Thus,  the
partition function in the grand-canonical ensemble is given by
\beq \ln{Z}=\ln\, {\rm det}(\dirac)\,. \label{particion} \eeq

In order to evaluate it in the zeta
 regularization  approach \cite{dowker}, we first determine the eigenfunctions,
 and the corresponding eigenvalues, of the Dirac operator, in the
 same gauge used in the previous section, i.e, we solve
\beq [-i\sigma_3(\partial_{\tau}+\mu)+
 i\sigma_2
\partial_x+\sigma_1
(i\partial_y-e B x)]\Psi=\omega \Psi\,,\eeq

To satisfy antiperiodic boundary conditions in the ${\tau}$
direction, we propose
  \beq
  \Psi_{k,l}({\tau},x,y)=\frac{e^{i\lambda_l {\tau}}e^{iky}}{\sqrt{2\pi\beta}}\,
  \psi_{k,l} (x)\,,\quad {\rm with}
  \quad
  \lambda_l=(2l+1)\frac{\pi}{\beta}\,,\label{lambda}\eeq
  where $\beta=\frac{1}{T}$ is the inverse temperature.

  After doing so, and writing $
\psi_{k,l} (x)=\left(\begin{array}{c}
  \varphi_{k,l} (x) \\
  \chi_{k,l} (x) \\
\end{array}\right) \,,
$ we have for each
  $k,l$, \beq \nn (\partial_x-k-e Bx)\chi_{k,l}&=&(\omega-\tilde \lambda_l)
\varphi_{k,l}
\\(-\partial_x-k-e Bx)\varphi_{k,l}&=&(\omega+\tilde \lambda_l) \chi_{k,l}
\,,\label{diffeq}\eeq where $\tilde \lambda_l=\lambda_l-i\mu$.

There are two types of eigenvalues

1. $\omega_l=\tilde \lambda_l$, with $l=-\infty,...,\infty$.

2. $\omega_{l,n}=\pm \sqrt{{\tilde \lambda_l}^2+2n e B}$, with
$n=1,...,\infty$, $l=-\infty,...,\infty$.

In all cases, the degeneracy per unit area is again given by
$\Delta_L$ in equation (\ref{deg}).

\section{Evaluation of the partition function at finite temperature and chemical potential}
\label{Sect3}

The partition function, in the zeta regularization scheme
\cite{elizalde}, is given by \beq \left.\log{{\cal
Z}}=-\frac{d}{ds}\right\rfloor_{s=0}\,\zeta(s,\frac{\dirac}{\alpha})\,.
\label{partfunc}\eeq As in the previous section, $\alpha$ is a
parameter with mass dimension, introduced to render the
$\zeta$-function dimensionless.

We must consider two contributions to $\log{{\cal Z}}$,
respectively coming from eigenvalues of type 1 and 2 in the
previous section, i.e., \beq \nn\Delta_1(\mu)&=&\left.
-\frac{d}{ds}\right\rfloor_{s=0}{\zeta}_1 (s,\mu)\\
&=&\left. -\frac{d}{ds}\right\rfloor_{s=0}\Delta_L \sum_{l=-\infty
}^{\infty}\left[ (2l+1)\frac{\pi}{\alpha
\beta}-i\frac{\mu}{\alpha}\right]^{-s}\,,\label{z1mu}\eeq and \beq
\fl \nn\Delta_2(\mu, B)&=&\left.
-\frac{d}{ds}\right\rfloor_{s=0}{\zeta}_2 (s,\mu, B)\\\fl
&=&\left. -\frac{d}{ds}\right\rfloor_{s=0}\!\!\!(1\!+\!(-1)^{-s})
\Delta_L \!\!\!\!\!\!\!\sum_{\begin{array}{c}
  n=1 \\
  l=-\infty \\
\end{array}}^{\infty}\!\!\!\!\!\!\!\left[\frac{2ne B}{\alpha^2}+
 {\left((2l+1)\frac{\pi}{\alpha
\beta}-i\frac{\mu}{\alpha}\right)}^2\right]^{-\frac{s}{2}}.\label{z2mu}\eeq

In the rest of this section, we sketch the main steps in the
analytic extension of both zeta functions and in the calculation
of their $s$-derivatives (for a detailed presentation, see
\cite{masslessnos}). The contribution $\Delta_1(\mu)$ can be
evaluated at once for the whole $\mu$-range. The analytic
extension of ${\zeta}_1 (s,\mu)$ can be achieved as follows (for a
similar calculation, see \cite{bagnos}) \beq \fl{\zeta}_1
(s,\mu)=\Delta_L \left(\frac{2\pi}{\alpha\beta }\right)^{-s}\left[
\zeta_H \left(s,\frac12-\frac{i\mu \beta}{2\pi}\right)+\sum_{l=0
}^{\infty}\left[
-(l+\frac12)-i\frac{\mu\beta}{2\pi}\right]^{-s}\right]\,.\eeq

Now, in order to write the second term as a Hurwitz zeta, we must
relate the eigenvalues with negative real part to those with
positive one without, in so doing, going through zeros in the
argument of the power. Otherwise stated, we must select a cut in
the complex $\omega$ plane \cite{ecz}. This requirement determines
a definite value of $(-1)^{-s}$, i.e., $(-1)^{-s}=e^{i\pi sign
(\mu)s}$. Taking this into account, we finally have \beq\fl
\zeta_1(s,\mu)=\Delta_L \left(\frac{2\pi}{\beta
\alpha}\right)^{-s}\left[\zeta_H \!\left(\!s,\frac12-\frac{i\mu
\beta}{2\pi}\right)+e^{i\pi sign(\mu)s}\zeta_H
\!\left(\!s,\frac12+\frac{i\mu \beta}{2\pi}\right)\right]
\label{ext1mu}.\eeq

From this last expression, the contribution $\Delta_1(\mu)$ to
$\log{{\cal Z}}$ can be obtained. It is given by \beq\fl\nn
\Delta_1(\mu) &=& -\Delta_L\left[\zeta_H^{\prime}\!\left(\!0,
\frac12- \frac{i\mu
\beta}{2\pi}\right)+\zeta_H^{\prime}\!\left(\!0, \frac12
+\frac{i\mu \beta}{2\pi}\right)+i\pi sign(\mu)\zeta_H \!\left(\!0,
\frac12 +\frac{i\mu \beta}{2\pi}\right)\right]\\\fl &=&\Delta_L
\left\{\log{\left(2\cosh{\left(\frac{\mu
\beta}{2}\right)}\right)}-\frac{|\mu|\beta}{2}\right\}\,.\label{delta1}\eeq

The analytic extension of ${\zeta}_2(s,\mu,B)$ requires a separate
consideration of different $\mu$ ranges. We study in detail two of
these ranges

\subsection{$ {\mu}^2 < 2e B$}

\beq \fl{\zeta}_2(s,\mu,B)=(1+(-1)^{-s})\, \Delta_L
\!\!\!\!\!\!\!\sum_{\begin{array}{c}
  n=1 \\
 l=-\infty \\
\end{array}}^{\infty}\!\!\!\!\left[\frac{2ne B}{\alpha^2}+
{\left((2l+1)\frac{\pi}{\alpha
\beta}-i\frac{\mu}{\alpha}\right)}^2\right]^{-\frac{s}{2}}\,.\label{z2mumenor}\eeq

Making use of the Mellin transform, this can be written as \beq
\fl \nn {\zeta}_2 (s,\mu,B)&=&\frac{(1+(-1)^{-s})\,
\Delta_L}{\Gamma(\frac{s}{2})}\sum_{n=1}^{\infty}\int_0^{\infty}dt\,
t^{\frac{s}{2}-1}e^{-t \left[\frac{2ne
B}{\alpha^2}+\left(\frac{\pi}{\alpha \beta}-\frac{i\mu}{\alpha
}\right)^2\right]}\\\fl&\times&\Theta_3\left(\frac{-2t}{\alpha
\beta}\left(\frac{\pi}{\alpha
\beta}-\frac{i\mu}{\alpha}\right),\frac{4\pi t}{(\alpha
\beta)^2}\right)\,,\eeq where we have used the definition of the
Jacobi theta function \hfil\break $\Theta_3
(z,x)=\sum_{l=-\infty}^{\infty}e^{-\pi x l^2} e^{2\pi z l}$.

To proceed, we use the inversion formula for the Jacobi function,
\hfil\break $ \Theta_3(z,x)=\frac{1}{\sqrt{x}}e^{(\frac{\pi
z^2}{x})}\Theta_3\left(\frac{z}{ix},\frac{1}{x}\right)$, and
perform the integration over $t$, thus getting \beq \fl\nn
{\zeta}_2(s,\mu,B)&=&\frac{(1+(-1)^{-s}) \Delta_L \alpha \beta
s}{4\sqrt{\pi}\Gamma(\frac{s+2}{2})}
\left[\Gamma\left(\frac{s-1}{2}\right)\left(\frac{2e B}{
\alpha^2}\right)^{\frac{1-s}{2}}\zeta_R
\left(\frac{s-1}{2}\right)\right.\\\fl
&+&\left.4\sum_{n,l=1}^{\infty}\!\!(-1)^l \left(\frac{l^2 \alpha^4
\beta^2}{8ne B}\right)^{\frac{s-1}{4}}\!\cosh{(\mu \beta
l)}\,K_{\frac{s-1}{2}}\left(\sqrt{2ne B\beta^2 l^2}\right)\right].
\label{ext2mu}\eeq

From this expression, the contribution $\Delta_2$ to the partition
function can be readily obtained, since the factor accompanying
$s$ is finite at $s=0$. After using that
$K_{-\frac{1}{2}}(x)=\sqrt{\frac{\pi}{2x}}e^{-x}$, and performing
the resulting sum over $l$, we obtain \beq \fl \nn
{\Delta}_2(\mu,B) &=&\Delta_L \beta \left[\sqrt{2e B}\zeta_R
\left(-\frac{1}{2}\right)\right.\\\fl
&+&\left.\frac{1}{\beta}\sum_{n=1}^{\infty} \log{\left(1+
e^{-2\sqrt{2ne B}\beta}+2\cosh{(\mu \beta )}e^{-\sqrt{2ne
B}\beta}\right)}\right]\,. \label{delta2}\eeq

Finally, adding the contributions given by equations
(\ref{delta1}) and (\ref{delta2}) we get, for the partition
function in the range ${\mu}^2 \leq 2e B$ \beq
\nn\log{Z}&=&\Delta_L \left\{\log{\left(2\cosh{\left(\frac{\mu
\beta}{2}\right)}\right)}-\frac{|\mu|\beta}{2}\right.+ \beta
\sqrt{2e B}\zeta_R \left(-\frac{1}{2}\right)\\&+&
\left.\sum_{n=1}^{\infty} \log{\left(1+ e^{-2\sqrt{2ne
B}\beta}+2\cosh{(\mu \beta )}e^{-\sqrt{2ne
B}\beta}\right)}\right\}\,.\label{zmenor}\eeq

\subsection{$2e B<{\mu}^2 < 4e B$}

As before, we have\beq \fl {\zeta}_2 (s,\mu,B)=(1+(-1)^{-s})\,
\Delta_L\!\!\!\!\! \sum_{\begin{array}{c}
  n=1 \\
  l=-\infty \\
\end{array}}^{\infty}\!\!\!\!\!\!\left[\frac{2ne B}{\alpha^2}+
{\left((2l+1)\frac{\pi}{\alpha
\beta}-i\frac{\mu}{\alpha}\right)}^2\right]^{-\frac{s}{2}}.\eeq

However, in this range of $\mu$, the contribution to the zeta
function due to $n=1$ must be analytically extended in a different
way. In fact, the expression cannot be written in terms of a
unique Mellin transform, since its real part is not always
positive (note, in connection with this that, for $n=1$, eq.
(\ref{ext2mu}) diverges). Instead, it can be written as a product
of two Mellin transforms \beq \fl \nn {\zeta}_2^{n=1}
(s,\mu,B)&=&\frac{(1+(-1)^{-s})}{\alpha^{-s}[\Gamma(\frac{s}{2})]^2}\,
\Delta_L \sum_{l=0 }^{\infty}\int_0^{\infty}dt\,
t^{\frac{s}{2}-1}e^{-\left[ (2l+1)\frac{\pi}{
\beta}-i\mu+i\sqrt{2e B}\right]t}\\\fl &\times&\int_0^{\infty}dz\,
z^{\frac{s}{2}-1}e^{-\left[ (2l+1)\frac{\pi}{
\beta}-i\mu-i\sqrt{2e B}\right]z}+\mu \rightarrow -\mu\eeq or,
after changing variables according to $t'=t-z; z'=t+z$, performing one of the integrals, and the sum over $l$ \beq \fl \nn
{\zeta}_2^{n=1}
(s,\mu,B)&=&\frac{(1+(-1)^{-s})\sqrt{\pi}}{2\alpha^{-s}\Gamma(\frac{s}{2})}\,
\Delta_L \left(2\sqrt{2e B}\right)^{\frac{1-s}{2}}\\\fl
&\times&\int_0^{\infty}dz\, z^{\frac{s-1}{2}}
J_{\frac{s-1}{2}}(\sqrt{2e B}z)\frac{e^{i\mu z}}{\sinh{(\frac{\pi
z}{\beta})}}+\mu \rightarrow -\mu \,.\eeq

Now, the integral in this expression diverges at $z=0$. In order
to isolate such divergence, we add and subtract the first term in
the series expansion of the Bessel function, thus getting the
following two pieces \beq \fl \nn {\zeta}_{2,(1)}^{n=1}
(s,\mu,B)&=&\frac{(1+(-1)^{-s})\sqrt{\pi}s}{4\alpha^{-s}\Gamma(\frac{s}{2}+1)}\,
\Delta_L \left(2\sqrt{2e B}\right)^{\frac{1-s}{2}}\\\fl\nn
&\times&\int_0^{\infty}dz\, z^{\frac{s-1}{2}}
\left[J_{\frac{s-1}{2}}(\sqrt{2e B}z)- \frac{\left(\frac{\sqrt{2e
B}z}{2}\right)^{\frac{s-1}{2}}}
{\Gamma\left(\frac{s+1}{2}\right)}\right]\frac{e^{i\mu
z}}{\sinh{(\frac{\pi z}{\beta})}}\\\fl &+& \mu \rightarrow
-\mu\,,\label{zeta21}\eeq and \beq \fl {\zeta}_{2,(2)}^{n=1}
(s,\mu,B)=\frac{(1+(-1)^{-s})\sqrt{\pi}}{2^s
\alpha^{-s}\Gamma(\frac{s}{2})\Gamma(\frac{s+1}{2})}\, \Delta_L
\int_0^{\infty}dz\, z^{{s-1}}\frac{e^{i\mu z}}{\sinh{(\frac{\pi
z}{\beta})}}+\mu \rightarrow -\mu \,.\label{zeta22}\eeq

The contribution of equation (\ref{zeta21}) to the partition
function can be easily evaluated by noticing that the factor
multiplying $s$ is finite at $s=0$. This gives
 \beq \nn {\Delta}_{2,(1)}^{n=1}(\mu,B)&=&\Delta_L\left\{\log{\left(1+e^{-2|\mu|\beta}
+2\cosh{(\sqrt{2e B}\beta)}e^{-|\mu|\beta}\right)}\right.\\
&+&\left.
|\mu|\beta -
2\log\left(2\cosh{(\frac{\mu\beta}{2}})\right)\right\}\,.\label{del21may}\eeq

In order to get the contribution coming from (\ref{zeta22}), the
integral can be evaluated for $\Re s>1$, which gives \beq \nn
{\zeta}_{2,(2)}^{n=1} (s,\mu,B)&=&\frac{(1+(-1)^{-s})\Gamma(s)
\sqrt{\pi}{(\alpha \beta)}^{s}
\Delta_L}{{(2\pi)}^{s}2^{s-1}\Gamma(\frac{s}{2})\Gamma(\frac{s+1}{2})}
\\&\times&\left[\zeta_H (s,\frac12
(1-\frac{i\mu\beta}{\pi}))+\zeta_H (s,\frac12
(1+\frac{i\mu\beta}{\pi}))\right] \,,\eeq where $\zeta_H(s,x)$ is
the Hurwitz zeta function. Its contribution to the partition
function can now be evaluated by using that $\zeta_H (0,\frac12
(1-\frac{i\mu\beta}{\pi}))+\zeta_H (0,\frac12
(1+\frac{i\mu\beta}{\pi})=0$ and the well known value of
$-\frac{d}{ds}\rfloor_{s=0}\zeta_H(s,x)$ \cite{gradshteyn}, to
obtain \beq
{\Delta}_{2,(2)}^{n=1}(\mu,B)=2\Delta_L\log(2\cosh{(\frac{\mu\beta}{2}}))
\,.\label{del22may}\eeq

Summing up the contributions in equations (\ref{delta1}),
(\ref{del21may}) and (\ref{del22may}), as well as the contribution
coming from $n\geq 2$, evaluated as in the previous subsection,
one gets for the partition function \beq \fl\nn \log{Z}&=&\Delta_L
\left\{\log{\left(2\cosh{\left(\frac{\mu
\beta}{2}\right)}\right)}+\frac{|\mu|\beta}{2}\right.\\ \fl &+&
\nn \log{\left(1+ e^{-2|\mu|\beta}+2\cosh{(\sqrt{2e
B}\beta)}e^{-|\mu|\beta}\right)} +\beta \sqrt{2e B}\left(\zeta_R
\!\left(\!-\frac{1}{2}\right)-1\right)\\\fl
&+&\left.\sum_{n=2}^{\infty} \log{\left(1+ e^{-2\sqrt{2ne
B}\beta}+2\cosh{(\mu \beta )}e^{-\sqrt{2ne
B}\beta}\right)}\right\} \label{zmayor}\,.\eeq

At first sight, this result looks different from the one
corresponding to $\mu^2<2e B$ (equation (\ref{zmenor})). However,
it is easy to see that both expressions coincide. The advantage of
using expression (\ref{zmayor}) for this range of $\mu$ is that
the zero-temperature limit is explicitly isolated from
finite-temperature corrections.

\section{Free energy and particle number}\label{Sect4}

 From equations (\ref{zmenor}) and (\ref{zmayor}), the free energy per
 unit area ($F=-\frac{1}{\beta}\log{Z}$) can be obtained
 \footnote{Consistently with the footnote in section \ref{Sect1}, all the results
 in this section are independent from the representation of the gamma matrices chosen.}. It is given
by \beq \fl\nn F(\mu, B, \beta)&=& -\Delta_L
\left\{\frac{1}{\beta}\log{\left(2\cosh{\left(\frac{\mu
\beta}{2}\right)}\right)}-\frac{|\mu|}{2}\right.+
 \sqrt{2e B}\zeta_R \left(-\frac{1}{2}\right)\\\fl &+&
\left.\frac{1}{\beta}\sum_{n=1}^{\infty} \log{\left(1+
e^{-2\sqrt{2ne B}\beta}+2\cosh{(\mu \beta )}e^{-\sqrt{2ne
B}\beta}\right)}\right\},\label{Fmenor}\eeq

Moreover, the free energy is continuous at ${\mu}^2 =2n e B,
n=0,...,\infty$. In the low-temperature limit one has \beq \nn
F({\mu}^2 < 2e B)\rightarrow_{\beta\rightarrow
\infty}-\Delta_L\sqrt{2e B}\zeta_R
\left(-\frac{1}{2}\right)\,,\eeq which coincides with the Casimir
energy obtained in section \ref{Sect1}, even for $\mu\neq0$, but
in this range, i.e., for $\mu$ less than the first Landau level,
if positive, or greater than minus the first Landau level, if
negative. On the other hand, \beq \fl \nn F(2e B<{\mu}^2 < 4e
B)\rightarrow_{\beta\rightarrow \infty}-\Delta_L \left\{\sqrt{2e
B}\left(\zeta_R
\left(-\frac{1}{2}\right)-1\right)+|\mu|\right\}\,.\eeq

The mean particle density can be obtained as
$N=\frac{1}{\beta}\frac{d}{d\mu}\log{Z}$. For nonzero temperature
and arbitrary $\mu$ (not coinciding with an energy level\footnote {Note that, for instance, if $\mu=2neB$, the series in equation (\ref{ext2mu}) converges only conditionally, and its term-by-term derivative leads to a divergent series.}) one has
\beq \nn N(\mu, B, \beta)&=&\Delta_L
\left\{\frac{1}{2}\left[\tanh{(\frac{\mu\beta}{2})}-sign(\mu)\right]\right.\\
&+&\left.\sum_{n=1}^{\infty} \frac{2\sinh{(\mu\beta)}e^{-\sqrt{2ne B}\beta}}{1+
e^{-2\sqrt{2ne B}\beta}+2\cosh{(\mu \beta
)}e^{-\sqrt{2ne B}\beta}}\right\}\,,\label{Nmenor}\eeq

It is interesting to note that, for $\mu=0$, one has
$N(\mu=0)=\pm\frac{\Delta_L}{2}$, which shows that, in the absence
of chemical potential, the fermion number obtained in equation
(\ref{nf}) remains unaltered as the temperature grows.

On the other hand, for nonvanishing $\mu$, the low-temperature
limit differs, depending on the $\mu$-range considered \beq \nn
N(2e Bn<{\mu}^2 < 2e B(n+1))\rightarrow_{\beta\rightarrow
\infty}n\Delta_L\, sign(\mu)\,,\eeq where
$n=\left[\frac{{\mu}^2}{2e B}\right]$.

This result is nothing but the expected one for particles with the
statistic of fermions, since relativistic field theory naturally
leads to the spin-statistics theorem. At zero temperature, $\mu$
is nothing but the Fermi energy; for example, for $\mu>0$, as
$\mu$ grows past a Landau level, such level becomes entirely
filled.

\section{Final comments}\label{Sect5}

From the previous result, the mean value of the particle density
at zero temperature can be obtained. After recovering units, one
has \beq \nn  j^0(2e c^2\hbar Bn<{\mu}^2 < 2e B
c^2\hbar(n+1))=\frac{-nce^2B}{h}\, sign(\mu)\,,\eeq the other two
components of the current density tri-vector being equal to zero
in the absence of an electric field.

Now, the zero-temperature limit of the same tri-vector in the
presence of crossed homogeneous electric ($F^{\prime}$) and
magnetic ($B^{\prime}$) fields can retrieved, for $F^{\prime}<c
B^{\prime}$, by performing a Lorentz boost with absolute value of
the velocity $\frac{F^{\prime}}{B^{\prime}}$. Suppose, for
definiteness, that the homogeneous electric field points along the
positive $y$ axis. Then, the velocity of the Lorentz boost must
point along the negative $x$-axis, and the transformation gives as
a result \beq \nn {j^{\prime}}^0=\frac{-nce^2B^{\prime}}{h}\,
sign(\mu)\,,\quad{j^{\prime}}^{x}=\frac{-ne^2F^{\prime}}{h}\,
sign(\mu)\,,\quad{j^{\prime}}^{y}=0\,.\eeq

As a consequence, the quantized zero-temperature Hall conductivity
is \beq \nn \sigma_{xy}= \frac{-ne^2}{h}\, sign(\mu)\,.\eeq

Finally, we mention that the more realistic case of massive
fermions is at present under study \cite{BGSS}.

\ack{We thank Paola Giacconi and Roberto Soldati for useful
discussions.\\ This work was partially supported by Universidad
Nacional de La Plata, under Grant 11/X381.}

\end{document}